\documentclass[conference]{IEEEtran}
\usepackage{float}
\usepackage{graphicx}
\usepackage{amsmath}
\usepackage{amsfonts}
\usepackage{amssymb}
\usepackage{setspace}
\usepackage{latexsym,subeqnarray,amsmath}

\begin{document}
%
\title{HPC Curriculum and Associated Ressources in the Academic Context}

\author{
\IEEEauthorblockN{Claude Tadonki}
\IEEEauthorblockA{
Mines ParisTech - PSL Research University \\
Centre de Recherche en Informatique (CRI)\\
35, rue Saint-Honor\'e, 77305, Fontainebleau Cedex (France)\\
Email: claude.tadonki@mines-paristech.fr}
}


\maketitle
\begin{abstract} Hardware support for high-performance computing (HPC) has so far been subject to significant advances. The pervasiveness of HPC systems, mainly made up with parallel computing units, makes it crucial to spread and vivify effective HPC curricula. Besides didactic considerations, it appears very important to implement HPC hardware infrastructures that will serves for practices, and also for scientific and industrial requests. The latter ensures a valuable connection with surrounding cutting-edge research activities in other topics ({\em life sciences, physics, data mining, applied mathematics, finance, quantitative economy, engineering sciences}, to name a few), and also with industrial entities and services providers from their requests related to HPC means and expertise. This aspect is very important as it makes an HPC Center becoming a social actor, while bringing real-life scenarios into the academic context. The current paper describes the major steps and objectives for a consistent HPC curriculum, with specific analyses of particular contexts; suggests how to technically set up operational HPC infrastructures; and discusses the connection with end-users, all these in both effective and prospective standpoints.
\end{abstract}


%
\IEEEpeerreviewmaketitle

\section{Introduction}
With the advent and pervasiveness of multicore processors, {\em parallel processing} is continuously getting out of mystery. However, even if it is true that most of end-users of common applications are now somehow aware of their parallel execution, and that ordinary programmers  are ready or more willing to admit the necessity of parallel processing considerations, delving into {\em parallel computing topics} \cite{book:1, book:3, book:4,book:16, book:14, book:15, book:17} is still not an instinctive nor intuitive step. People need to be convinced about that choice, and the balance between the underlying effort and the corresponding reward should appear favorable. Thus the need of incentive HPC initiatives.

Significant efforts have been made so far to make {\em parallel computing} implementation as easy as possible, if not seamless from the programming viewpoint \cite{book:13, book:7, book:8, book:10, book:11}. However, the topic is still non trivial \cite{book:21, book:19}. To date, there is no effective compiler who can automatically derive a parallel version of a given ordinary sequential code without explicit user annotations or monitoring. Indeed, {\em parallel computing} adds another abstraction to {\em basic algorithm design} that has nothing to do with the underlying problem being solved. Thus, the topic might be perceived as a plain complication, otherwise technically speculative. Thus, we need to grab people attention and make them accept to take the way of {\em parallelism}. This is an education goal, prior to  targeting an increasing number of (potential) candidates for parallel computing curriculum.

There is a increasing need of programmers with parallel programming skills. Thus, it is now common to see job ads seeking people qualified or specialists in {\em parallel computing}. This has significantly motivated {\em parallel computing curricula}, which, from our point of view, should not stand totally disconnected from other areas. Indeed, it is always better to understand what we have to parallelize. This will result in a more skillful parallel scheme and yield a consistent self-contained report. Thus, it is preferable to have good basic skills before moving to parallelism. This through has to be kept in mind while designing an HPC curriculum.

From a technical point of view, especially with {\em shared memory parallelism} \cite{book:9}, there is a heavy hardware concurrency, which should be managed or taken into account carefully at design or evaluation time  \cite{book:6}. Thus, it is important to make sure that the {\em basis of computer architecture} \cite{arch1, arch2} are well understood, maybe making it a prerequisite. Another reason for this is the impact that a parallel scheme might have on {\em numerical aspects}. Also for {\em proof of correctness},  hardware mechanisms of parallel systems should be included into the equation. Unfortunately, it is very common to encounter {\em computer architecture unaware} parallelizations. This might be due to a lack of related skills or just a plain  carelessness.  The most noticeable example is the important proportion of programmers who are completely unaware of {\em vector computing}, also referred as {\em SIMD computing}, whose related hardware elements and mechanisms are implemented in almost all modern processors. Thus, providing a computer architecture course at the earliest stage is certainly a good idea. 

We see that there are numerous facts and thoughts that emerge when it comes to {\em parallel computing} from a didactic viewpoint. Thus, we need to efficiently organize and manage all related  teaching/training/expertise activities \cite{tad}. Beside invariant common needs, there are specific contextual issues that need to be addressed conscientiously through well targeted applications. Within an academic context, the best way to handle all these initiatives is to set up an HPC Center that will implement the main didactic actions and promote the use of {\em parallel computing} as previously stated. The purpose of this paper is to provide a consistent panoramic view of an academic HPC Center including the key topics with their interaction, and to describe its main associated  activities.  The rest of the paper is organized as follows. The next section describes key didactic factors related to HPC, followed by an overview of how to design and implemented an HPC Center. Section IV addresses the question of building a local HPC cluster. Section V concludes the paper.

\section{Didactic key factors about HPC}
\subsection{Justify why considering parallel computing}
First of all, we have to sell the need for HPC \cite{hpc0, hpc1, hpc2} to our (potential) audience. This preliminary communication is crucial, because people need to be convinced that the related investment is worth considering. The requested effort is not only technical, but also conceptual, as it implies a change of mind about computing devices, programs, algorithms and programming. This should be a clear winning step in order to capture a sincere and keen attention of the listeners, their enthusiasm will determine the success of the course and associated exchanges. We should always keep in mind that ordinary people expect the increasing CPU frequency to fit their needs for speed. The way to achieve the aforementioned marketing goal is to raise a number of (intentionally {\em pro domo}) arguments. We state some of them:
\begin{itemize}
\item
There are numerous important real-life applications where processing speed is essential, but which cannot be fulfilled at the expected level by a single CPU \cite{book:20,hpcapp4,tad6}. Nice and illustrative examples should be described here, some of them following an anecdotal way if possible. Complicated cases should be avoided, even if (and this can be at most mentioned) {\em scientific computing} is the major ``customer'' of parallel computing \cite{book:5, book:12, book:13, tad2}. We suggest for instance: {\em urgent query in a large distributed database}; {\em real-time high definition image processing} \cite{tad5}; {\em realistic high precision simulation}\cite{tad1, tad3}; {\em new generation video gaming}. With a scientific audience, we use to display figure Fig.\ref{run} with {\em difficult network graphs problems} as examples.
\begin{figure}[h]
\centering
\includegraphics[scale=1]{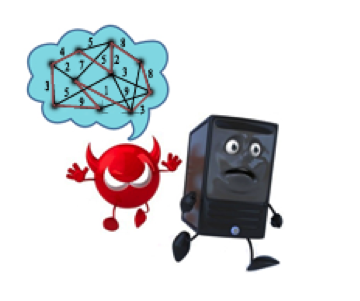}
\caption{\label{run}Computing need standing over computers capability}
\end{figure}
\item
The so-called {\em Big Data} is an active topic for good reasons. Indeed, the volume of data to be processed in number of real-life applications is more and more huge, and the response is expected to remain almost instantaneous whatever the scenario. The case of social networks is a good example here. Data analytics applications can be also mentioned. The major actors of the two previous topics are developing a significant research activity in parallel computing, with explicit laboratories implemented in various countries.    
\item
Almost all modern processors integrate several CPU cores. This fact by itself can also be considered as a plain argument for delving into parallel computing concepts. Even end-users of applications that run in parallel could get a better technical picture from being aware of that. Personally, we use to introduce the current fact by considering the OS {\em multitasking}, which allows to concurrently run several applications even with a single CPU core. Then explain that this virtual concomitance becomes effective with parallel units.
\item
Energy is a major concern in most of modern activities including those related to the use of computers. It should be recall here that energy is nearly proportional to running time \cite{tad4}. Thus, reducing the energy consumed by a given software application can be achieved by accelerating its execution.  In addition,  a processor with clock frequency $f$ consumes more energy than a dual-core with frequency $f/2$ per core. The slogan here is ``{\em parallel computing is energy efficient}''. 
\item
Many natural activities are intrinsically parallel. Thus, {\em natural parallelism} is intuitively expected to be so implemented on computers. Illustrative examples can be taken from {\em simulation} and {\em  artificial intelligence}.
\end{itemize}    
\subsection{Keep HPC useful for others}
This is very important in order to make the HPC Center being perceived as a useful entity and thereby receive rewarding considerations. In addition, keep addressing external requests related to {\em industrial activities}, {\em scientific research}, {\em societal issues},{\em services}, and {\em trainings},  will certainly confirm the importance of the Center and vivify its activities. In practice, it is common to see financial sponsors firmly recommend such connection and also use it a crucial metric for their evaluation and decision. The same exhortation  might come from institutional hierarchy or politics. Note that, because of the strong need for HPC availability, some companies implement their own HPC section. Academics could claim permanent investigations, less restricted research focus, and high potential for innovation.
\subsection{Make sure to address local needs}
From our personal teaching and collaboration experience, in South America for instance, it clearly appeared that addressing local needs and issues is highly appreciated and valuated. This is normal because otherwise, especially in case of too specific concerns, there is no chance that an appropriate solution will come from elsewhere. In addition, there is a strong desire to have direct hand to the implemented solutions for many reasons including {\em cost}, {\em suitability}, {\em independence}, and {\em technical maturity}. Thus, an HPC Center should prioritize local issues in their applications portfolio, of course with a certain of level interest in external and prospective scenarios.
\subsection{Be keen in academic collaborations}
Standing and acting in an isolated manner is a noticeable factor of stagnation and sort of obscurantism, which is the opposite of what is expected from an academic entity. HPC is moving too fast and changes occur very frequently. Hardware advances and near future projections are regularly announced. New HPC methods, achievements, and challenges are actively made available. Thus, keep collaborating with other actors is essential to survive and to receive consideration. The resulting synergy is another vital benefit to take into account.
\section{Implementing an academic HPC Center}
This section discusses the main components and objectives of an academic HPC Center.
\subsection{Global organization of the Center}
\begin{figure*}
\centering
\includegraphics[scale=0.6]{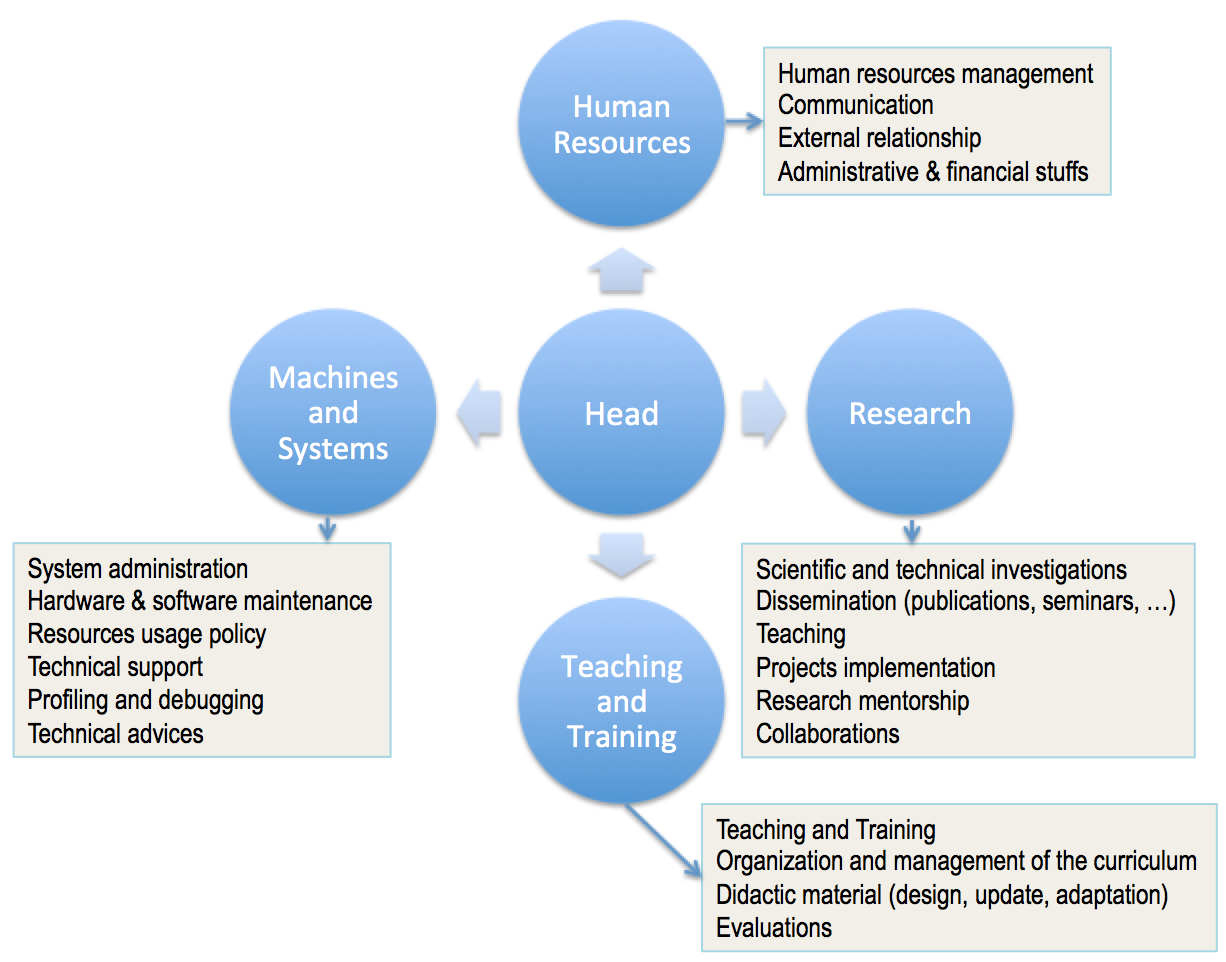}
\caption{\label{Center}Overview organization of an academic HPC Center}
\end{figure*}

The Center itself is a virtual entity of people with relevant skills. Figure Fig. \ref{Center} provides a view of the main units of the Center. Beside this logical clustering, it is important to have the units interacting efficiently. External requests as previously explained are likely to reach the head or research staff. A management policy should (at least roughly) indicate how process such requests.  
\subsection{A consistent HPC curriculum}
We present and discuss the main basis for a consistent HPC curriculum \cite{tad}. Still at the level of an overview, we provide a connected panorama of the key learning units related to {\em parallel computing}. The teaching part could be complemented with external invited speakers, thus the importance of collaborations, which will also boost the outcome of the research activities.
\begin{figure*}
\centering
\includegraphics[scale=0.62]{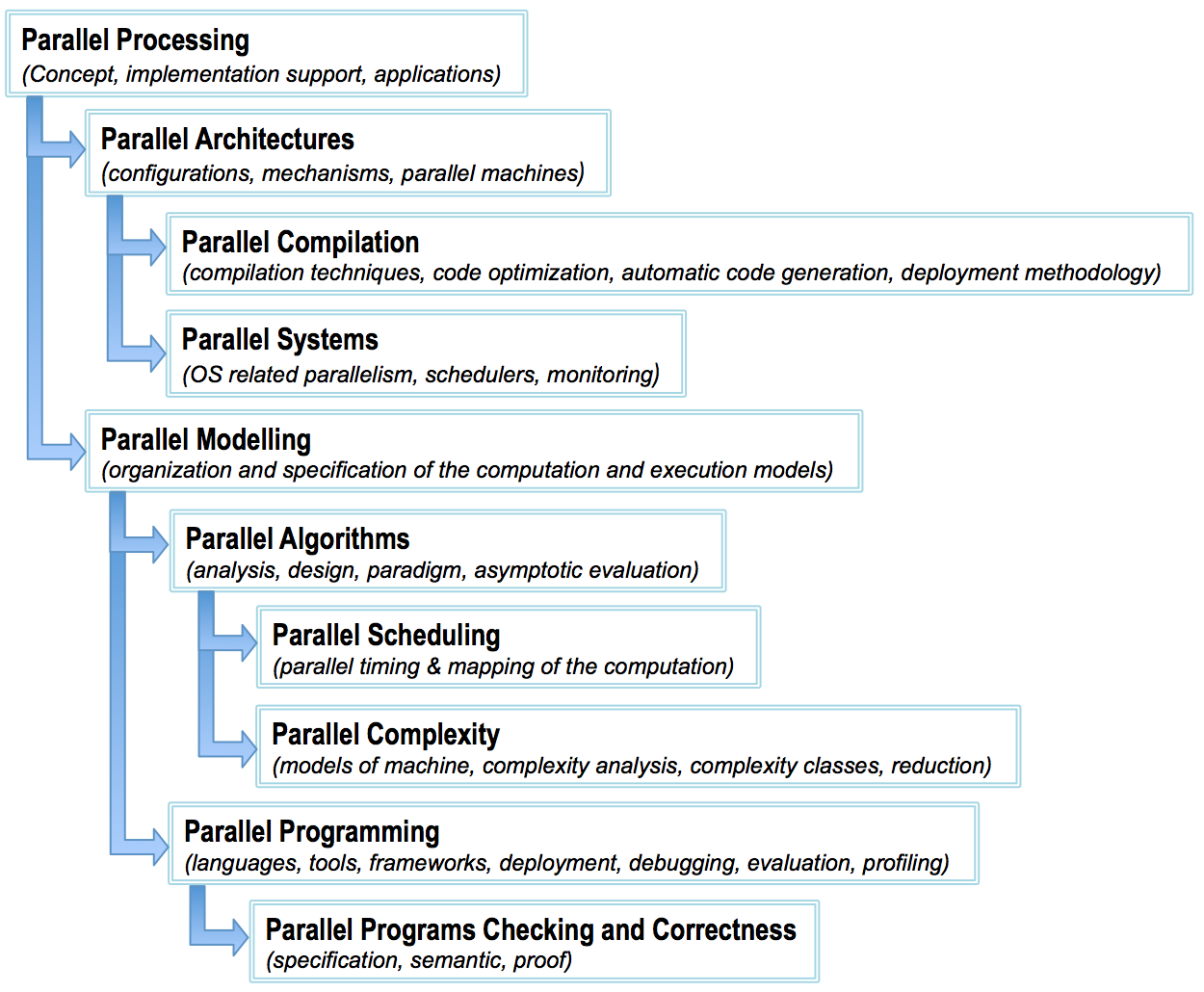}
\caption{\label{dep}Dependencies within a parallel computing curriculum}
\end{figure*}
\subsubsection{Topics selection}
\begin{figure*}
\centering
\includegraphics[scale=0.2]{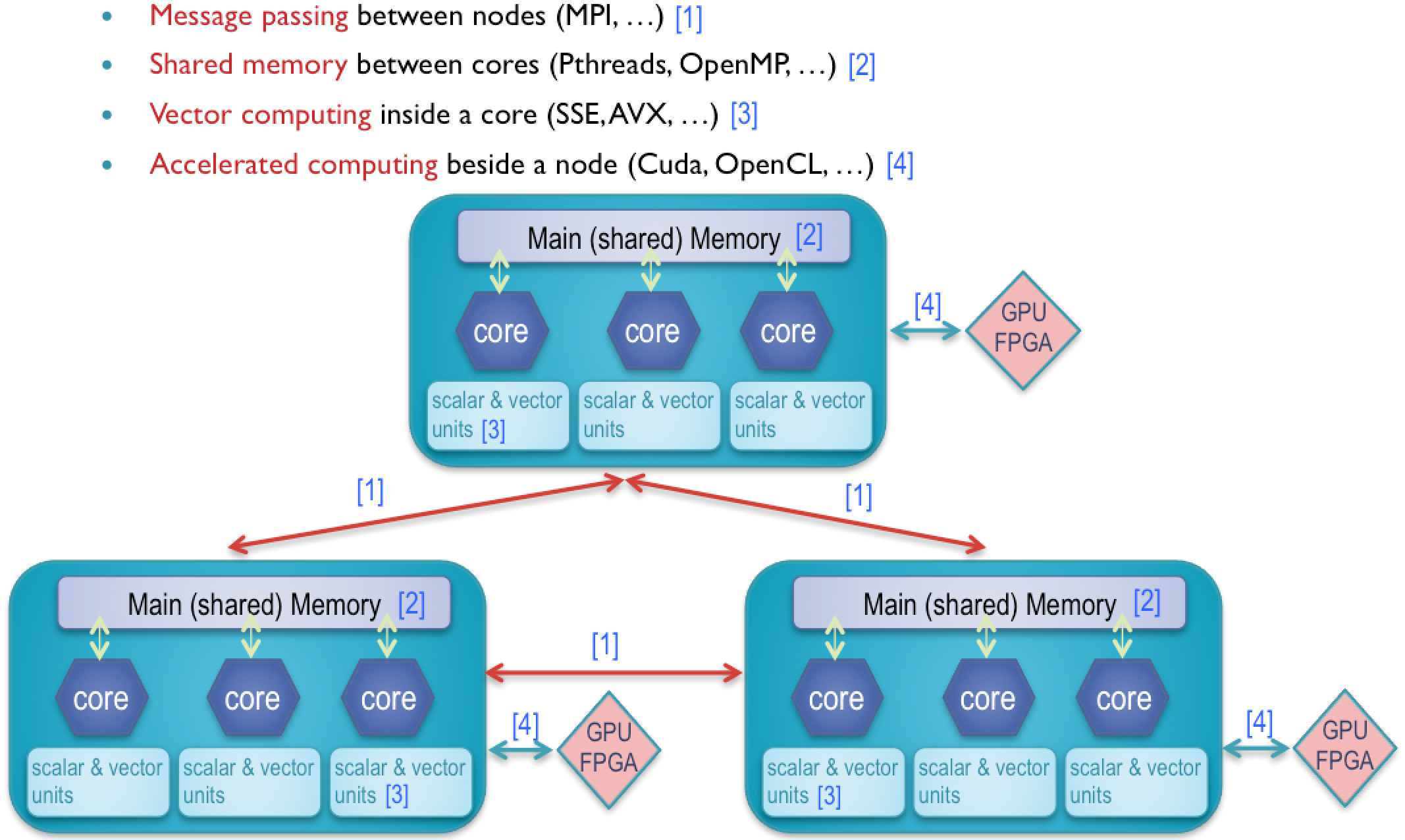}
\caption{\label{axes}Main paradigms of parallel programming }
\end{figure*}

\begin{figure*}
\centering
\includegraphics[scale=0.6]{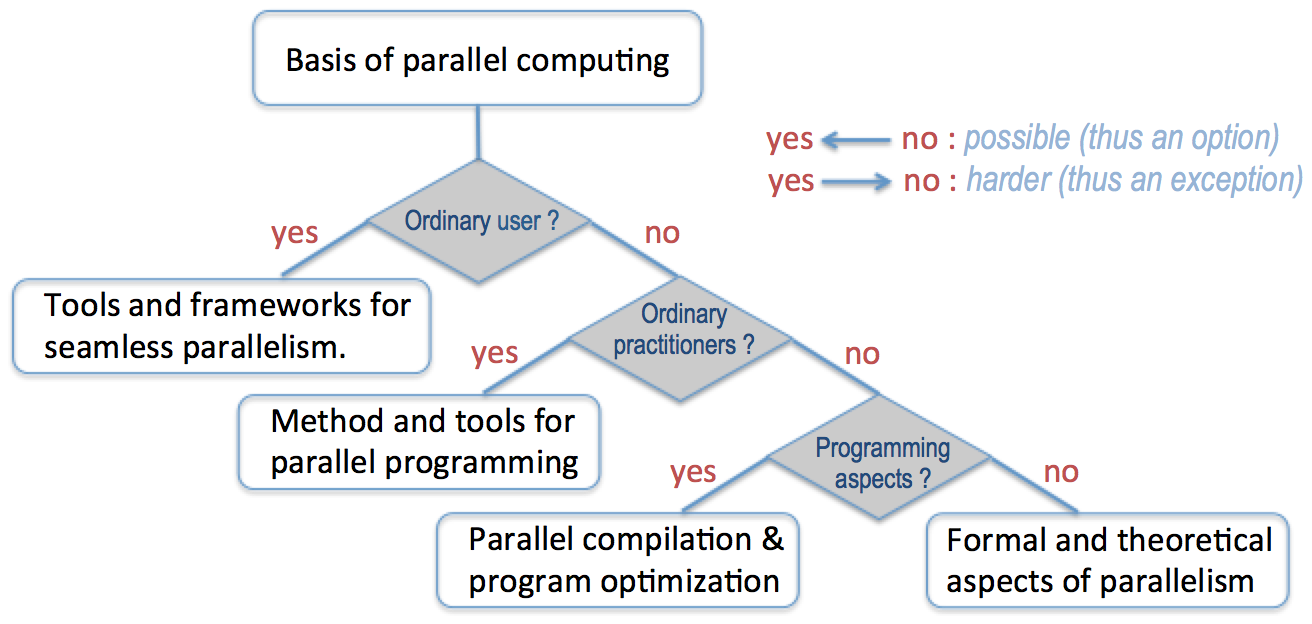}
\caption{\label{chart}Flowchart of parallel computing curriculum}
\end{figure*}

The curriculum should be organized so as to allow coherent inner steps and motivated choices. Figure Fig. \ref{axes} shows the main paradigms of {\em parallel programming}. These paradigms are independent, so the associated classes can be provided without any strong interconnection. However, it should be made possible to follow all of them for those seeking a multi skilled profile, which is technically essential for {\em hybrid parallel programming}.  
Regarding how to chose the right way,  figure Fig. \ref{chart} suggests a a basis for a systematic selection.
\subsubsection{Applications selection}
As previously explained, the selection of the applications to be used as case studies is very important in order to best capture the attention of the audience and also create a strong connection with reality. The effect of parallel computing should convincing. Thus, the main criteria for selecting the target applications are: {\em not purely fictive}, {\em good illustrative potential}, {\em high impact}, {\em well parallelizable}, {\em not to complicated to implement in parallel}. For instance, Bioinformatics topics are particularly exciting in the Brazilian context. Indeed, Brazil has a great biodiversity, however knowledge about the genetic resources and the biotechnological potential of Brazilian species is still very limited. Bioinformatics motivations, challenge and perspectives as they are stated by UnB researchers include: 
\begin{itemize}
 \item 
Great diversity of organisms in Brazil with biotechnological potential
 \item 
Major infectious diseases among Brazilian population have been the focus of intensive research
 \item 
Number of genome and transcriptome projects is increasing very fast due to Brazilian government support
 \item 
Huge amount of data has been generated and has to be processed and analyzed
 \item 
Need for active scientific collaborations to provide an efficient computing support to genomic research, with an emphasis on pipeline optimization, information storage and recovering. 
\end{itemize}

We now list some applications extracted from our past and current collaborations:
\begin{itemize}
 \item 
 One of the main problems Bioinformatics is facing now is how to process genomic data at the rate it is being produced. Once a new biological sequence is discovered, its functional/structural characteristics must be esta\-blished. For this purpose, newly discovered sequence is compared against the sequences that compose genomic databases, in search of similarities. In order to accelerate the production of results, we could use parallel/distributed variants of the exact methods for pairwise or sequence-profile comparisons, employing many computing units in order to reduce the computation time.
 \item
 Flight scheduling is one of the earliest steps in airline operations
planning. The schedules quality is crucial for airline profitability and operations reliability. The {\em fleet assignment} problem consists in deciding which aircraft types (or fleet) will be assigned to the flights. This problem is routinely considered in airline industry, thus the need of a fast computer program to reach good solutions in a reasonable time. 
 \item 
 One of the biggest challenge in Life Science is reading and analyzing the genetic code of the genome. Current high-throughput sequencing machines are able to produce hundreds of millions of short sequences (reads) in a single run and the cost of sequencing  has decreased dramatically.  However, the task of reconstructing genomes from the large volumes of fragmentary read data still remains an algorithmic challenge that is in practice answered by heuristic and non-satisfactory solutions. Exact methods are computationally costly, thus the need of parallel computing.
\item
The Amazon has a long history of human settlement. The knowledge that Amazonians have about natural plants and their properties are being lost over time. Currently, there are some attempts to store these information by creating documented catalogs of plants and further analyze them. In this context, number of south America countries that share the Amazonian territory joint together to help Amazonians making their ecosystem safer. The amount of collected information is huge, and process them efficiently is best done by parallel computing. 
\end{itemize}

\section{Setting up an operational HPC cluster}
Any connected machine can be accessed remotely through the network. This is the case for most HPC clusters for which such an access is commonly made available for external users. This is also the mean by which HPC resources are shared between  different entities. Nevertheless, there are several reasons for planning a local HPC cluster: 
\begin{enumerate}
\item
fulfilling routine computing requests from local entities; 
\item
allowing internal or private computing activities; 
\item
serving as a support for parallel computing practices; 
\item
acting as a resources center; 
\item
getting or claiming a computing resources independence; 
\item
fulfilling the need or reaching a milestone of an HPC project; 
\item
housing a multi-purpose server; 
\end{enumerate}
to name a few. We now describe the way to target this technical achievement and discuss the main items.

The first technical question, maybe implicit, should be ``{\em what is the total peak performance that we want for the supercomputer}?'', followed by practical concerns like {\em design budget}, {\em maintenance budget}, {\em lifetime}, {\em jobs density and frequency}, {\em cooling requirement}, to name a few. Other secondary questions might focus on {\em the vendor}, {\em specific hardware capabilities}, {\em hardware flexibility}, {\em robustness}, {\em suitability to the expected usage}. From the answers of these questions, a picture of the global configuration can be drawn. Now, let move to technical aspect considerations regarding the goal of building a proprietary HPC cluster.

\begin{figure}[h]
\centering
\includegraphics[scale=0.7]{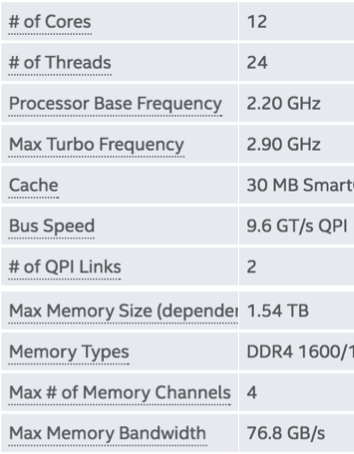}
\caption{\label{brd}Intel Broadwell-EP main characteristics}
\end{figure}

The first and major concern should be the generic compute node, which could be a {\em multicore CPU}, a {\em mono-socket manycore CPU}, or a {\em multi-socket manycore CPU}, coupled or not with GPU card(s). An example of compute node could be a \{mono/multi\}-socket Intel Broadwell-EP \cite{broadwell}. Figure Fig.\ref{brd} displays its main specifications. 
There pros and cons for aggregating to many cores within a single node.  The main advantage is the one usually considered for hybrid share/distributed memory computing, which is the reduction of the interconnection network. Having less physical nodes is also an acceptable argument. Another one, very important, is energy, both heat and power consumption. Concerning the cons, the situation is like putting all your eggs within the same basket. In case of (even local) failure, it is likely to disable (and loose) the whole node, which thus yields a severe consequence. Another one is at the OS side, user threads (in a multithreaded program) will collude with standard processes. From the sustained performance viewpoint, a heavy concurrency between too many threads threads within the same node might severely affect the program scalability. This bottleneck mainly results from penalizing memory accesses, which is a crucial concern with NUMA configurations. 

Next, since the available compute nodes will have to communicate to each other, hence come questions related to the network like {\em speed (bandwidth and latency)}, {\em topology}, and {\em reliability}. Considering a simple Ethernet network will be sufficient and operational. 

Storage capacity should be evaluated and extended if necessary. Also for this item, backup mechanisms should be addressed. 

For software aspects, the choice will focus on {\em operating system}, {\em compilers}, {\em programming and runtime libraries}, {\em debug tools}, {\em monitoring tools}, and {\em (batch) schedulers}.

For a technically detailed step-by-step explanation and tutorial to build up a standard HPC cluster, we suggest \cite{build}.

\section{Conclusion}
Designing and implementing a parallel computing curriculum has became a common goal. Indeed, the topic of {\em parallel and distributed computing} is now vital because of the emergence and pervasiveness of parallel systems. This topic has several directions and a wide range of applications, thus should be considered a good level of conscience. Managing all related activities within an HPC Center is clearly a good idea, which needs to be considered methodologically. The enthusiasm behind is motivated by the active evolution of HPC systems and the increasing demand of efficient computing solutions for numerous applications.  Important facts, advices, schematic views, recommendations and descriptions are reported in this paper in the perspective of creating an active  academic HPC Center.

\end{document}